# IN SEARCH FOR SIGNS OF CHAOS IN BRANCHING PROCESSES

**Zhen CAO and Rudolph C. HWA**

Institute of Theoretical Science and Department of Physics
University of Oregon, Eugene, Oregon 97403-5203

**Abstract**

For systems that involve particle production through branching processes the concept of chaos is explored. The measures that can describe their behaviors are investigated. Monte Carlo simulation is used to generate events according to perturbative QCD and an Abelian model. It is shown how the measures proposed distinguish the two cases in ways that characterize the chaotic behavior.

It has been known for some time that the nonlinear, non-Abelian dynamics of the classical Yang-Mills field has chaotic solutions [1, 2]. More recently, it has been shown by lattice calculation that the classical non-Abelian gauge theory generally exhibits deterministic chaos and that the Lyapunov exponents can be numerically determined [3, 4, 5]. How to extend the investigation to the quantum theory is, however, unclear inasmuch as the notion of quantum chaos for such dynamics is not well defined [6, 7]. In this paper we take the first step in that direction, not just going into the quantum dynamics of a nonlinear theory, but into the realm of particle production of quantized fields.

In this uncharted territory we have very little guidance on what to study in search for signs of chaos. It is not clear what a trajectory is in QCD, even less the distance between two trajectories. What exactly is the Kolmogorov entropy, well defined in classical dynamics [8], is also unclear in the multiparticle production problem. Our first objective is therefore to the find some measure that can play the role of distance between trajectories and some other quantity that conveys the loss of information in the final state.

In order to know whether or not we have found the right measures, it is necessary to test our ideas on different dynamical problems and show that the measures can effectively distinguish different characteristics. We shall therefore concentrate on two branching processes in particle production. One is the pure gluon theory in perturbative QCD; the other is a cascade model (to be referred to as the $\chi$ model) that has none of the features of the gauge theory. We shall regard the gluons (partons) as particles and ignore hadronization so that we may focus on the issue of chaoticity in the branching dynamics.

In either case we start with a parton having high time-like virtuality $Q^2$ and use computer simulation to study the evolution of the system through branching. The Monte Carlo code generates the parton momentum distributions that in the case of QCD satisfy the Altarelli-Parisi evolution equations [9, 10, 11]. The splitting functions for the two cases are drastically different, and the ways the virtualities degrade are also different. They give rise to the diverse behaviors that emerge. The question is how to find a quantitative measure of the diversity that is useful from the point of view of chaoticity.

One of the difficulties in the problem posed is that time is not an explicit variable when calculating momentum distribution. Moreover, with the number of degrees of freedom increasing in the cascade process, there is no obvious way to generalize the definition of a trajectory in classical dynamics. However, branching has an irrefutable physical notion of the direction of time, and parton multiplicity $n$ increases with time so long as only tree diagrams (without recombination) are considered, which we do. Thus we take two conjugate views: the "temporal" development (parameterized by $n$) without regard to the parton momenta, and the "geometrical" characteristics in the momentum space at the end of the evolution. In some rough sense this corresponds to the two views of the classical systems that are related by ergodicity.

Concerning trajectory and distance between trajectories, our thinking goes as follows. The classical field intensity is replaced in the quantum case by the number $n$ of quanta, and the distance between two field configurations can be represented by the variance $D^2$ of the $n$ distribution, $P_n$. This distribution arises after repeated simulation from the same initial condition, i.e., a fixed initial virtuality $Q^2$. Quantum fluctuation is enough to replace the small variation of the initial condition in classical dynamics. With the average $\langle n \rangle$ regarded as a measure of the evolution time, $D^2$ as a function of $\langle n \rangle$ can then be regarded as the quantum analog of the classical distance function $d(t)$. This $\langle n \rangle$ is the average multiplicity, not just in the final state, but along the evolution process.

More precisely, let us focus on one event and use a tree to represent a particular branching process. Regardless of the virtuality $q^2$ of any line, all vertices of the same generation are put at the same branching level and a label $i$ is given to each generation between two levels, starting with $i = 0$ for the initial parton at $Q^2$. By ignoring $q^2$ in the genealogy of the tree, we are emphasizing topology and overlooking momenta. Let $n_i$ denote the number of partons at the $i$th generation. We shall use the event averaged $\langle n_i \rangle$ in lieu of time, although a simple linear or exponential relationship between the two is not expected. In fact, $\langle n_i \rangle$ may not increase monotonically with $i$, even if $n_i$ does; in such a case we use only the increasing portion. For the measure that plays the role of distance, we define the normalized variance

$$V_i = \left( \left\langle n_i^2 \right\rangle - \langle n_i \rangle^2 \right) / \langle n_i \rangle^2 \tag{1}$$

This differs from the second cumulant moment $K_2$ by a term $-1/\langle n_i \rangle$, and is better than it because $K_2$ can become negative (thus precluding log-log plots), while $V_i$ is always positive. A rapid increase of $V_i$ with $\langle n_i \rangle$ can be interpreted as the analog of the "divergence of nearby trajectories" in classical dynamics. Clearly, if $V_i$ remains constant or decreases, one does not become more ignorant about the state of the system as it evolves, and thus it cannot be regarded as exhibiting chaotic behavior.

For MC simulation we follow the procedure described by Odorico [10]. For both pure-gauge QCD and the $\chi$ model we start with virtuality $Q^2$ and end at $Q_0^2$. The splitting function for $g \to gg$ is

$$P(z) = 6 \left[ \frac{1-z}{z} + \frac{z}{1-z} + z\,(1-z) \right] \qquad \text{(for QCD)} \tag{2}$$

where $z$ is the momentum fraction of the daughter parton in the frame where the mother parton's momentum is 1. In terms of the Sudakov form factor $S(Q^2, Q_0^2)$ one can calculate the probability of emitting a resolvable gluon occurring between $Q^2$ and $Q_0^2$; when it occurs in the simulation at $q^2$, then $z$ is chosen to be in the range $z_0 \leq z \leq 1 - z_0$, where $z_0 = Q_0^2/q^2$. The daugthers evolve separately from the maximum virtualities, $q_1^2 = z\,q^2$ and $q_2^2 = (1-z)\,q^2$, according to the same procedure as applied to the mother. The running coupling constant is, as usual, $\alpha_s(q^2) = 4\pi/[11 \log(q^2/\Lambda^2)]$, where we have set $N_c = 3$, $N_f = 0$, and shall use $\Lambda = 250$ MeV. Branching ceases when $q^2$ reaches $Q_0^2$ or below.

In the $\chi$ model we use the splitting function

$$P(z) = 6z\,(1-z) \qquad \text{(for the } \chi \text{ Model)} \tag{3}$$

There is no infrared or collinear divergence and therefore no evolution. Nevertheless, we introduce $Q^2$ dependence by requiring that the daughter virtualities be $zq^2$ and $(1-z)q^2$, when the mother virtuality is $q^2$, and $z$ can be any value between 0 and 1. We require branching to occur successively until the virtualities of all lines reach $Q_0^2$. We consider this model because it examplifies the Abelian dynamics without infrared and collinear divergences. Yet the multiplicity of particles produced depends on $Q^2$ so that the result of branching can be compared with the non-Abelian case, although for very different $Q/Q_0$.

We have simulated these two branching processes by running $10^5$ events each, using $Q_0 = 1$ GeV. The results on $V_i$ vs $\langle n_i \rangle$ are shown in the log-log plot in Figure 1 for various values of $Q/Q_0$ indicated. While it is hard to produce high multiplicity in QCD unless $Q/Q_0$ is extremely large, particles are copiously produced in the $\chi$ model at moderate $Q/Q_0$. The general features of $V_i$ vs $\langle n_i \rangle$ for the two cases are markedly different. The QCD result shows a power-law increase in the high $\langle n_i \rangle$ range

$$V_i \propto \langle n_i \rangle^{\kappa} \quad , \qquad \kappa \simeq 0.4 \quad , \tag{4}$$

where the exponent $\kappa$ is approximately independent of $Q^2$. The $\chi$ model, on the other hand, shows a rapid rise initially, but followed by a precipitous drop after reaching a maximum. Indeed, the maximum $V_i$ decreases with increasing $Q^2$. Clearly, this is not a case that suggest chaotic behavior.

For gluon branching the monotonic increase of $V_i$ with $\langle n_i \rangle$ implies that, as the branching proceeds, how many particles are produced in any event becomes more and more unpredictable. The power-law dependence may be regarded as the analog of the exponential increase with time of the distance between classical trajectories that are initially close by. However, there is no way to relate $\kappa$ to the Lyapunov exponents, since among other differences the notion of time is not well defined here. By itself $\kappa \simeq 0.4$ does not indicate how chaotic the behavior is. There is a need for another measure of chaoticity.

If Figure 1 is viewed as the analog of the description of the temporal behavior, another place to search for signs of chaos is in the phase space of the particles. As the system evolves, more and more information is lost on where the partons are (or more precisely, what their momenta are), so entropy increases not only because of the increase of the number of particles, but also because of the dynamical fluctuations in their momenta. We therefore consider a multifractal description of that fluctuation and focus on the information dimension as a characterization of the entropy of the system [8].

Since at each vertex of branching a daughter's momentum fraction $z$ is known in the simulation, the momentum $x$ of a final particle as a fraction of the initial particle is therefore calculable. It is $x = \prod_i z_i$, where $z_i$ is the momentum fraction of the descendant at the $i$th generation. Since the particle distribution $\rho(x)$ is highly peaked near $x = 0$, it is smoother to examine the distribution in the cumulative variable $X$, defined by [12]

$$X(x) = \int_{x_1}^{x} \rho(x')\, dx' / \int_{x_1}^{x_2} \rho(x')\, dx' \quad , \tag{5}$$

where $x_1$ and $x_2$ are two extreme points in the distribution $\rho(x)$, between which $X$ varies from 0 to 1. In terms of $X$ the inclusive distribution $\rho(X)$ is constant. For each event the fluctuation in $X$ space is then studied by dividing the interval $0 \le X \le 1$ into $M$ bins and calculating the factorial moments

$$f_q(M) = M^{-1} \sum_{j=1}^{M} n_j (n_j - 1) \cdots (n_j - q + 1) \tag{6}$$

where $n_j$ is the multiplicity in the $j$th bin. After averaging over all events, the normalized factorial moment

$$F_q = \langle f_q \rangle \,/\, \langle f_1 \rangle^q \tag{7}$$

is known to contain no statistical fluctuations [13].

What interests us is its behavior near $q = 1$, where we can extract the information dimension $D_1$. To that end it is necessary to extend the definition of $F_q$ in (6) to noninteger $q$. A method for achieving that without losing the attribute $F_q = 1$ for Poissonian fluctuation has recently been developed [14]. Using that method we have calculated $F_q$ for a continuous range of $q$, as shown in Figure 2, for both QCD and the $\chi$ model. The results for the two cases are very different and provide a distinct contrast between them. They are in accord with the temporal behavior shown in Figure 1 in that, for $q > 1$, $F_q < 1$ in the $\chi$ model, meaning that the distribution is sub-Poissonian, while $F_q > 1$ in QCD, indicating large fluctuations. The geometrical properties in the $X$ space are not revealed until we investigate the $M$ dependence. We find that in both cases $F_q$ are not sensitive to $M$, as can be seen from the various lines in Figure 2 corresponding to different $M$ values. Thus the behavior has no interesting multifractal property: $D_1 \simeq 1$ in both cases.

The origin of the lack of significant $M$ dependence can be traced to $F_q$ itself, where the event averaging cancels out the fluctuations. Event by event the values of $F_q^e = f_q^e/(f_1^e)^q$, where $e$ labels an event, fluctuate greatly, especially when $M$ is large. To quantify the degree of that fluctuation we define event-averaged (vertical) moments of the (horizontal) $F_q^e$ moments

$$C_{p,q}(M) = \left\langle F_q^p(M) \right\rangle / \left\langle F_q(M) \right\rangle^p , \tag{8}$$

where $\left\langle F_q^p \right\rangle = \mathcal{N}^{-1} \sum_e (F_q^e)^p$, $\mathcal{N}$ being the total number of events. We then calculate $C_{p,q}(M)$ for $0 < p < 2$, and $q = 2, 3, 4$. It is found that $C_{p,q}(M)$ indeed exhibits significant $M$ dependences, as shown in Figure 3. In all cases of $p$ and $q$, the $\chi$ model has smaller $C_{p,q}$ compared to QCD, implying smaller fluctuations of $F_q^e$. The $M$ dependences do not show linearity over any extended range in the log-log plots, the best being from $M = 5$ to 20. In that range we write

$$C_{p,q}(M) \propto M^{\psi_q(p)} \quad . \tag{9}$$

From the slope $\psi_q(p)$ in the neighborhood of $p = 1$, we can calculate the index $\mu_q$, defined by

$$\mu_q = \frac{d}{dp} \psi_q(p) \left.\right|_{p=1} \quad . \tag{10}$$

The result of our simulation yields the values $\mu_q = 0.0061, 0.054, 0.23$ for $q = 2, 3, 4$ in the case of QCD, and 0.0014, 0.010, 0.046 in the $\chi$ model, respectively. Clearly, for each $q$, $\mu_q^{(\mathrm{QCD})}$ is significantly larger than $\mu_q^{(\chi)}$. We now give a physical interpretation of $\mu_q$ as an entropy index to be used as a new measure of the event fluctuation in branching processes.

If we define $P_q^e = F_q^e / \sum_e F_q^e$, and then define $H_{p,q} = \sum_e (P_q^e)^p$, we have an entropy in the event space

$$S_q = -\sum_e P_q^e \ln P_q^e = -\frac{d}{dp} \ln H_p \,|_{p=1} \quad . \tag{11}$$

$H_{p,q}$ and $C_{p,q}$ can be related by their definitions, yielding

$$\frac{d}{dp} \ln C_{p,q} \,|_{p=1} = \ln \mathcal{N} - S_q \quad . \tag{12}$$

From (9) and (10) we obtain

$$S_q = \ln(\mathcal{N} M^{-\mu_q}) \quad , \tag{13}$$

apart from a possible additive term independent of $\mathcal{N}$ and $M$. $\mu_q$ appears to be related to the information dimension, but it is not because $S_q$ is not the entropy defined in the $X$ space, which is the one that is divided into $M$ cells. The event space in which $S_q$ is defined has not been partitioned into small cells. The meaning of (13) can be seen in two extreme cases: (a) if $F_q^e$ is the same for every event, then $P_q^e = 1/\mathcal{N}$, and $S_q = \ln\mathcal{N}$; (b) if only one event has $F_q^e \neq 0$, and $F_q^e = 0$ in all others, then $S_q = 0$. Thus case (b) is more ordered in the event space than (a); that is, it is more disordered to spread out an observable ($F_q^e$ here) over all events (even if $F_q^e =$ constant) than to confine it to a few events having nonzero values (analogous to the increase of entropy of an expanding gas). [The case of all $F_q^e = 0$ is excluded from consideration in order to render $P_q^e$ meaningful.] Thus $S_q$ *decreases* when there are *more* events with $F_q^e = 0$, signifying more order in the event space. From (13) we see that $\mu_q$ is a measure that decrease, which in turn implies more fluctuation in $F_q^e$.

At large $M$ only large spikes in small bins contribute to $F_q^e$, especially when $q$ is large. Events with large spikes are rare. Consequently, the fluctuation in $F_q^e$ from event to event becomes more pronounced with increasing $q$. That behavior is now quantified by $\mu_q$. We may therefore use $\mu_q$ to characterize the spatial properties of the chaotic behavior of a branching process. We have, however, at this stage no quantitative criterion on how *small* $\mu_q$ must be in order to signify *no* chaotic behavior.

We can relate the classical and quantum problems in our description in the event space as follows. Consider a neighborhood $N_\epsilon$ around an initial point in phase space for a classical trajectory. For chaotic dynamics, starting the system from any point in $N_\epsilon$ leads to widely different trajectories at sufficiently long time later. We may regard $\mathcal{N}$ of these trajectories in $N_\epsilon$ as corresponding to $\mathcal{N}$ branching events all with the same initial virtuality, but having different outcomes. $F_q^e(M)$ describes the final state of the system for the $e$th event, and $\mu_q$ describes the degree of fluctuation of $F_q^e$ from event to event. Sufficiently large values for the index $\mu_q$ therefore signify chaotic behavior of the branching process.

In conclusion, we have found features about QCD branching that are not shared by the $\chi$ model, which represents Abelian branching. Because of the non-classical nature of the system, we have had to search for new measures and observables. The dependence of $V_i$ on $\langle n_i \rangle$ reveals the "temporal" behavior, while $F_q$, $C_{p,q}$, and $\mu_q$ describe the "spatial" properties. All these measures taken together give a collective description of the degree of chaoticity in a branching process. We have found that $V_i$ increases with $\langle n_i \rangle$ in QCD, while it decreases for the $\chi$ model. The dependences of $F_q$ on $q$ are totally different for the two cases. $C_{p,q}$ and $\mu_q$ are much larger for QCD than for the $\chi$ model. These results collectively suggest that QCD branching is chaotic, while the $\chi$ model is not. Among the measures considered, $V_i$ (and higher moments of $n_i$, which could also have been considered) vs $\langle n_i \rangle$ contain detailed information about the branching process from generation to generation, but they are not accessible to the experiment. The others describe the characteristics of the final state, and can be determined experimentally in most high-energy collisions. The entropy index $\mu_q$ is most unusual and deserves further investigation.

One of us (RCH) is grateful to K. Geiger, B. L. Hao, S. G. Matinyan, and B. Müller for helpful discussions. This work was supported in part by the U.S. Department of Energy under Grant No. DE-FG06-91ER40637.

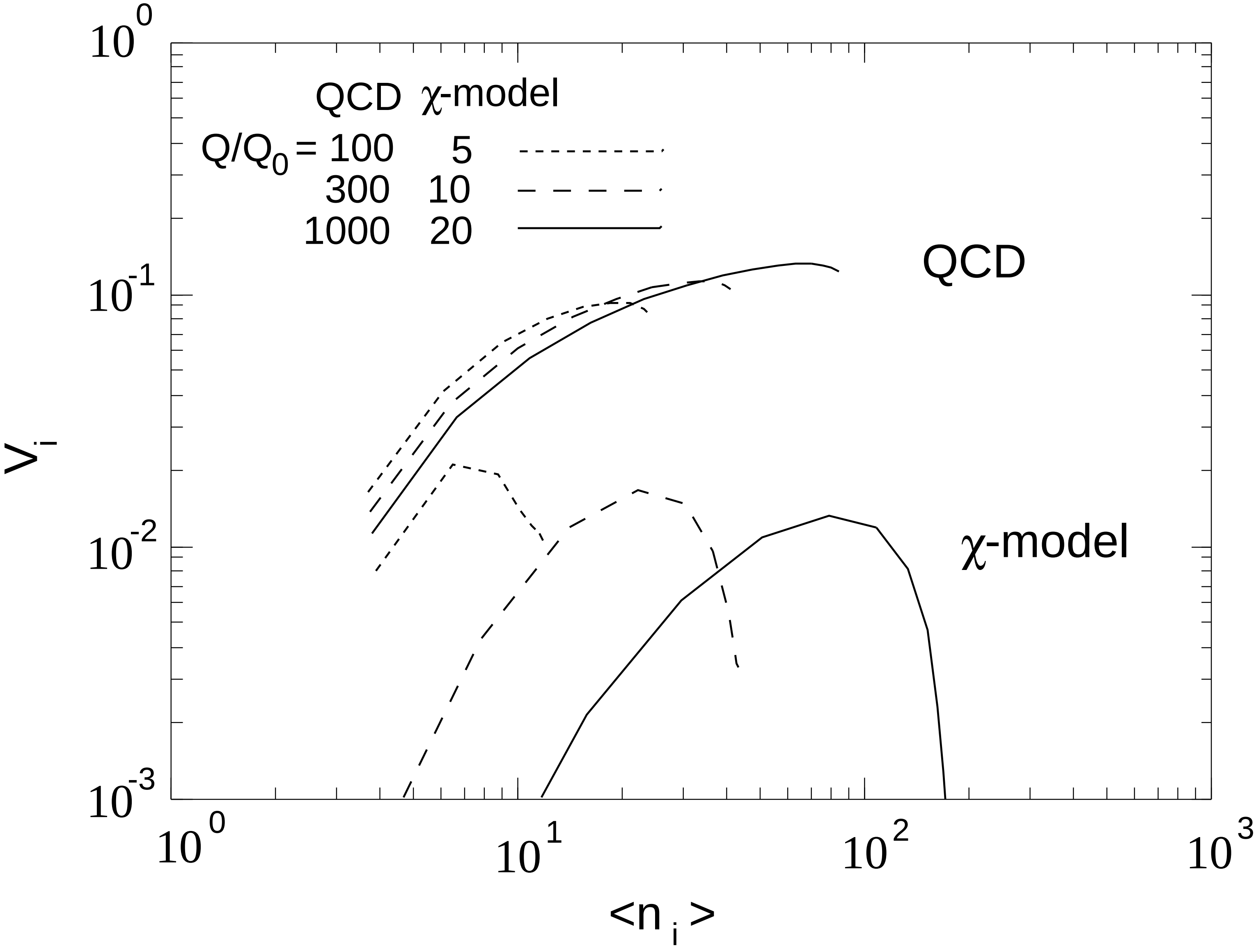

Figure 1: Normalized variance vs average multiplicity at various generations in the branching processes.

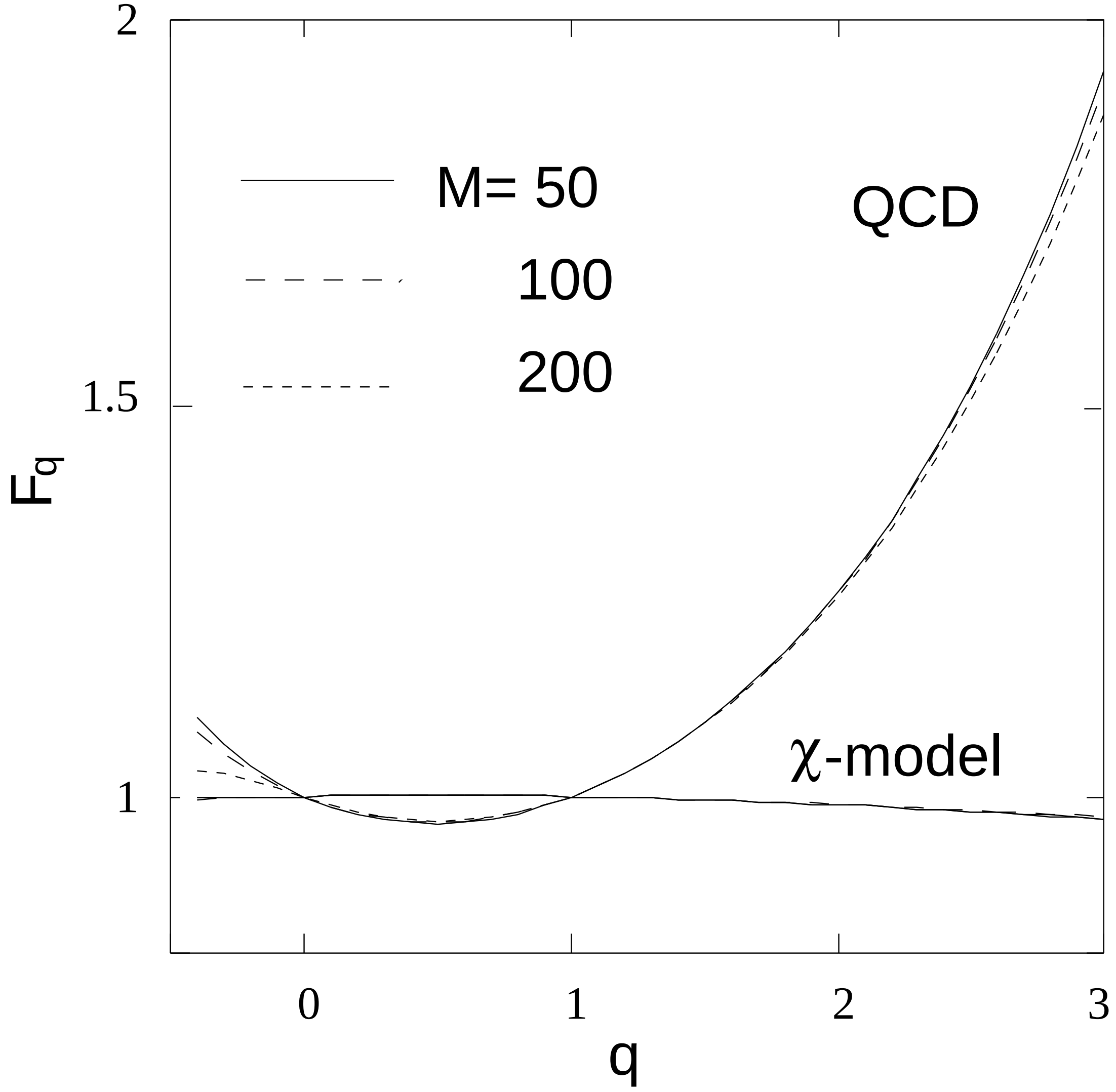

Figure 2: Factorial moments of continuous order for various bin sizes. $Q/Q_0 = 10^3$ for QCD and 20 for the $\chi$-model.

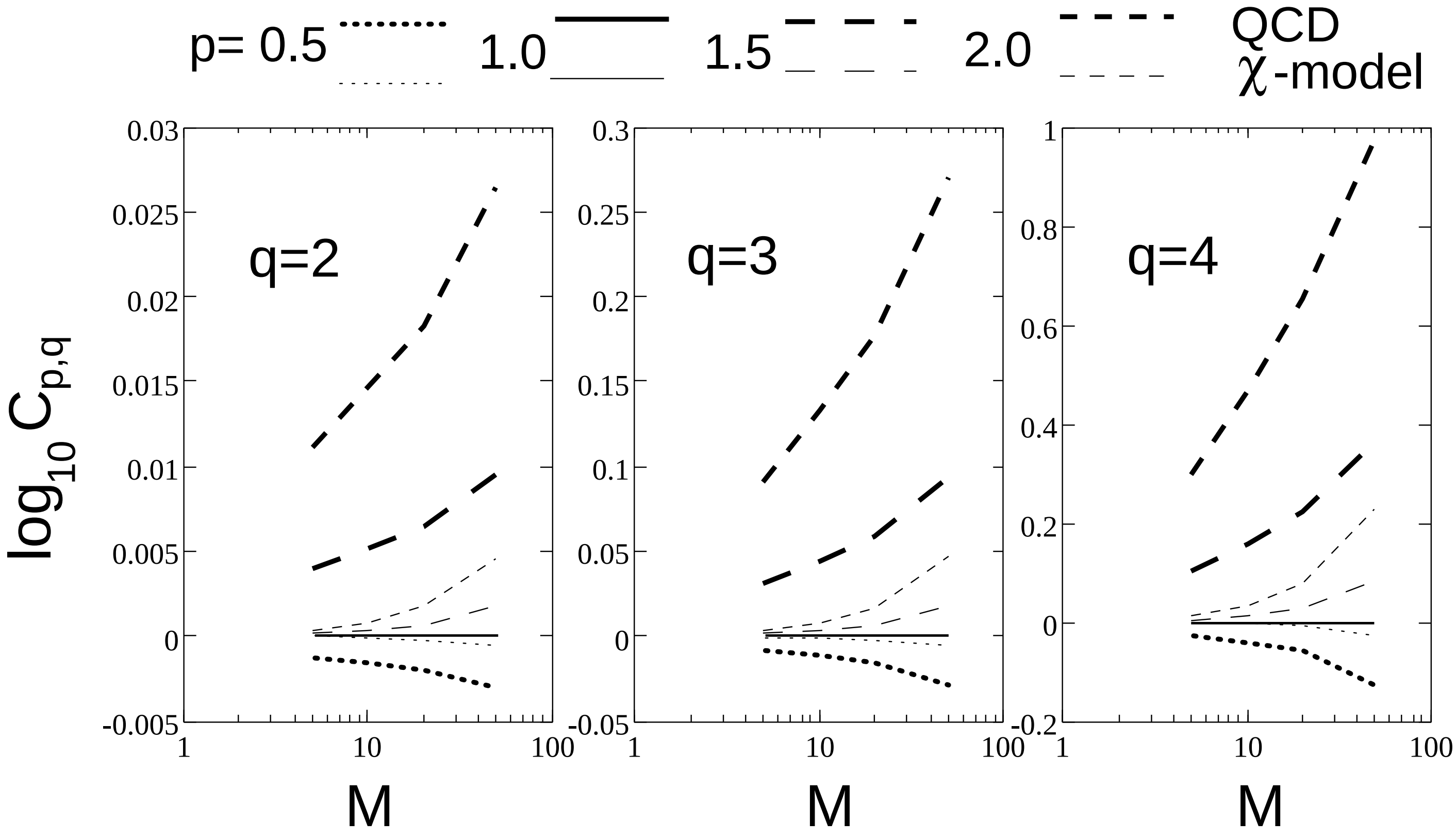

Figure 3: $C_{p,q}$ vs $M$ for various values of $p$ and $q$. $Q/Q_0 = 10^3$ for QCD and 20 for the $\chi$-model.